# Micro-optics Fabrication using Blurred Tomography

**Daniel Webber,**[1,*] **Yujie Zhang**[1], **Kathleen L. Sampson**[1], **Michel Picard**[1], **Thomas Lacelle**[1], **Chantal Paquet,**[1] **Jonathan Boisvert**[1], **and Antony Orth**[1]

[1]*National Research Council of Canada, Ottawa, Ontario, Canada K1A 0R6*

**daniel.webber@nrc-cnrc.gc.ca*

**Abstract:** We demonstrate the fabrication of millimeter-sized optical components using tomographic volumetric additive manufacturing (VAM). By purposely blurring the writing beams through the use of a large etendue source, the layer-like artifacts called striations are eliminated enabling the rapid and direct fabrication of smooth surfaces. We call this method blurred tomography, and demonstrate its capability by printing a plano-convex optical lens with comparable imaging performance to that of a commercially-available glass lens. Furthermore, due to the intrinsic freeform design nature of VAM, we demonstrate the double-sided fabrication of a biconvex microlens array, and for the first time demonstrate overprinting of a lens onto an optical fiber using this printing modality. This approach to VAM will pave the way for low-cost, rapid-prototyping of freeform optical components.

*Introduction*

Additive manufacturing (AM) has transformed the way optical components are created. Optical elements made with AM benefit from the intrinsic freeform design nature of the process, enabling the fabrication of custom components not possible using conventional manufacturing techniques. [1-9,12,14] Additive manufacturing of optical components have been demonstrated using several technologies. One such technique is inkjet printing where the sequential deposition of micro-droplets form the final part [1,2]. Other techniques include vat-polymerization techniques such as digital light processing (DLP) [3,4] and stereolithography (SLA) [5,6] where the optical component is fabricated in a layer-by-layer fashion, two-photon polymerization (2PP) where microscopic volumes of photoresin are sequentially solidified to form the final part [7,8,9], and more recently in volumetric additive manufacturing (VAM) where the entire part is printed simultaneously using tomographic means [12,14].

In all but the latter of these techniques, parts are printed onto a planar substrate and material is built up either one voxel (FDM, SLA, 2PP), line (Inkjet), or layer (DLP) at a time. As a result of the processes used, parts created using these techniques face several issues: 1. Part surfaces that are not parallel to the substrate have steps with height defined by the material layer thickness, 2. Parts with curvature on multiple sides (e.g. Biconvex lens) or suspended lens structures are difficult to fabricate with SLA, DLP, or Inkjet as they require support structures, and 3. Part fabrication is slow as the material must be built sequentially (0D, 1D, or 2D) to construct the final 3D structure.

In this paper, we overcome these issues using VAM to fabricate imaging-quality lenses. VAM represents a paradigm shift in additive manufacturing. Compared to other AM techniques that build parts in a layer-by-layer fashion, VAM builds all regions of the part simultaneously, avoiding layer-artifacts and enabling the production of smooth surfaces. Furthermore, VAM can fabricate complex structures without the need for support structures, and is rapid since all regions of the printed part are formed simultaneously.

Despite not being a layer-based method, VAM suffers from layer-like artifacts (commonly called striations) within the vial plane that lead to ridges on the part surface [10,11,12], prohibiting the creation of imaging-quality lenses. It was shown in Rackson et al. [13] that these striations are caused by a self-writing waveguide effect. During the final stages of printing, the narrow writing beams used in VAM causes accelerated printing in planes parallel to the writing beams. By terminating the print prior to the onset of photoresin gelation, and then flood illuminating the entire print volume, Rackson et al. were able to mitigate striations since the source is not pixelated.

Other post-processing techniques have been implemented to achieve smooth surfaces in VAM and the production of optically-smooth surfaces. Using a silica-embedded polymer, Toombs et al. [14] were able to fabricate glass microlenses with sub-10 nm surface roughness. This was achieved by first printing the object using the silica-polymer matrix forming a "Green Part". Next, the part is baked in an oven at 600°C to debind the polymer from the silica forming the "Brown Part". Finally, the part is sintered at 1300°C to melt the silica nanoparticles into a solid fully dense part. In a more recent work, Peng et al. [12] fabricated polymer-based plano-convex optical elements with sub-nanometer RMS surface roughness by using a meniscus-coating method; a thin-film of uncured photoresin is allowed to remain on the as-printed object, and through wetting and capillary action, filling in any imperfections in the part surface forming a round and smooth surface.

All of the methods described above require varying degrees of additional post-processing steps to achieve the final part. In the latent imaging approach [13], precise knowledge of photoresin polymerization and subsequent flood exposure time is required to obtain the final part. In Toombs et al. [14], a multistep part baking process is required to achieve fused silica glass adding both complexity and time, counteracting the rapid-prototyping advantage VAM offers. The method of Peng et al. [12] relies on a thin layer of uncured resin to remain on the lens surface. As a result, the surface is susceptible to damage from handling, and may deform if exposed to a light source with wavelength in the absorption spectrum of the photoresin.

A direct VAM method that does not require additional post-processing steps would be ideal for maintaining the rapid-prototyping nature without the need for added processing variables and equipment, while at the same time maintain the freeform design nature of additive manufacturing.

Here, we demonstrate that preferentially blurring of the optical writing beam in volumetric printing enables the fabrication of surfaces with sub-nm surface roughness which is essential for the production of optical components. Unlike typical volumetric printers with well-collimated, low etendue writing beams that satisfy the conditions of the Filtered Backprojection method [15], or software-based techniques to correct for light scattering [16] and dose blurring [17] effects, we instead purposely introduce blurring to the beam via the use of a large etendue projection source. Coupled with the astigmatism introduced by the cylindrical vial without index matching bath [17,18,19,20], blurring is achieved across the entire print volume. Unlike previous techniques that relied on intricate post-processing [12,14] or additional printing steps [13], the method described here can directly fabricate smooth surfaces for any arbitrary object orientation in the print volume using simple post-processing methods.

## 2. Methods

### 2.1 Blurred Tomographic Printer for Lens Fabrication

In this work, a tomographic additive manufacturing printer was used to fabricate micro-optical components. Fig. 1 (a) shows a system diagram of the printer. A digital light innovations CEL5500 UV projector (DMD pitch 10.8 μm) with f/# = 4.68 telecentric lens system transmits tomographic projections of the voxelized target object onto an 8 mm outer-diameter vial containing photoresin. The vial is mounted to a Physik Instrumente M-060 rotation stage. Due to this arrangement, the pixel size at the focal plane of the projector is 10.8 μm with etendue 4.153E-6 sr mm$^2$. The vial of photoresin functions as a cylindrical lens which introduces astigmatism to the projector field causing points within the vial plane and along the vial axis to be focused at different points along the optics axis of the projector. This can be readily seen in images of the photoresin fluorescence taken from overhead the vial (Fig. 1(b)) and from the side of the vial (Fig. 1(c)). Furthermore, due to the finite etendue of the optical source, we observe a foreshortening of the depth of field as observed by the sharp focusing of the writing beams.

Astigmatism and blurring of the writing beam can be visualized in Fig. 1(d). Here, the dose delivered by a downsampled projector pixel was simulated using 3DRT, and shown for planes at different depths within the vial. Typically in VAM, the side length of an individual voxel is defined by the projector pixel size, however, due to the computational complexity of 3DRT, the voxel space representing the print volume was downsampled by a factor of four, yielding a voxel size (and effective pixel size in our calculations) of 43.2 μm. Cross-sectional slices of the planes at -1.5 mm, 0 mm, and 1.5 mm along the vial axis (Z) and vial plane (X) are shown in Fig. 1(e). At the vial center (Y = 0 mm), the beam profile is much wider than the beam width in the vial plane. As we move towards the rear of the vial (Y = 1.5 mm), we observe a decrease in width along the vial axis and subsequently an elongation along the vial plane. Typically, this blurring is undesirable as it reduces the spatial resolution of the printing process [10] and introduces asymmetry to the dose delivery [17,18]. However, for printing objects with features larger than the point spread of the blur, it is preferential as it eliminates striations in the final print, enabling the production of smooth surfaces.

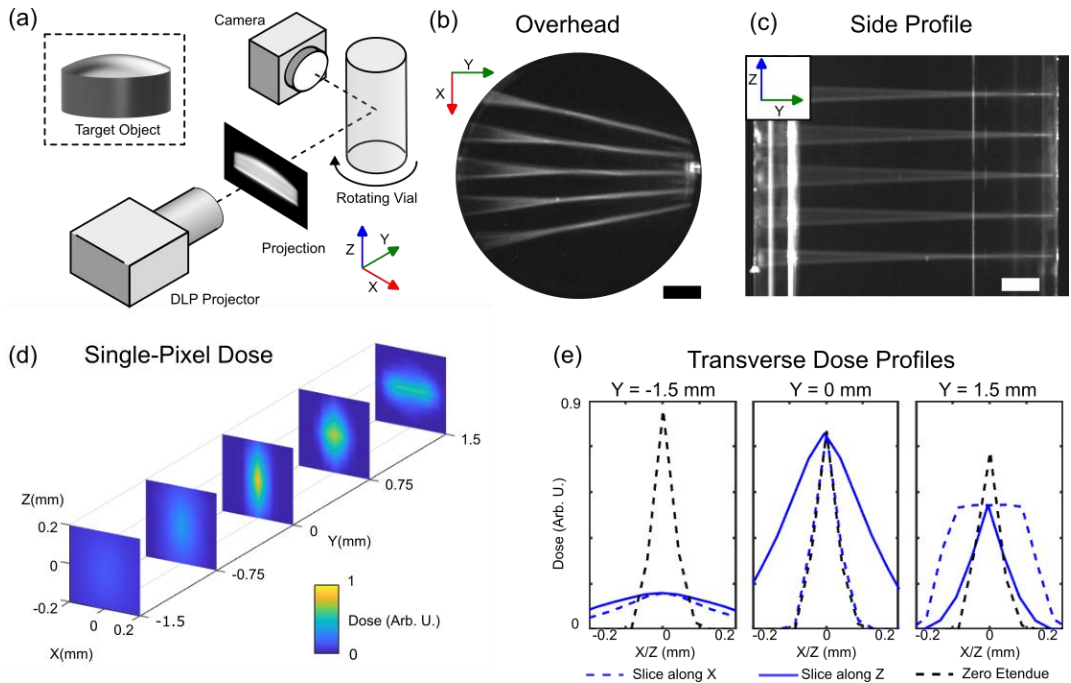

Figure 1: Creating smooth surfaces with blurred tomography. (a) Diagram of setup. Here, a custom projection lens with adjustable aperture stop enables printing with large etendue. Images of photoresin fluorescence show both astigmatism and depth of field foreshortening in the print volume, as measured from above the vial (b) and from the side (c). (d) Numerical calculations of the dose delivered by a single pixel on the optical axis of the projector. (e) Transverse dose profiles of the numerical calculations in (d) at three different depth planes, as well as for the ideal case of zero etendue. Scale bars in (b,c) are 1 mm.

### 2.2 Tomographic projection calculation

Due to the foreshortened depth of field of the writing beams caused by the large etendue conditions, the light dose delivered to the print volume will not match the target design resulting in dimensional errors in the printed part. To compensate for this error, we use a technique called 3D ray-tracing (3DRT) that we introduced in a previous report [18] to compute the tomographic projections. For each pixel in the projector, a cone of rays (cone angle set by the etendue of the system) are transmitted through the optical system of the printer into the print volume. The coordinates of these rays are then used to first estimate a tomographic projection of the target design, and then based on that projection compute the light dose delivered to the print volume. This process is iterated to optimize the delivered dose to match the target design. Because cast rays span the area of the system aperture, etendue related effects can be compensated for in the tomographic projection optimization process. For the prints shown here, 20 iterations of the optimization process were performed, and 19 rays were cast per-pixel to model etendue in the printing system.

### 2.3 Post-processing of printed lenses

After printing, parts were removed from the vial and suspended in isopropyl alcohol (Sigma Aldrich 99.5 %) for 1 minute to remove most of the uncured photoresin, followed by an additional 20 minutes in a second isopropyl alcohol bath to remove any residual photoresin. Afterwards, they were removed and placed on a microscope slide and left to air-dry at room temperature until all solvent was visibly removed. Finally, parts were placed in a vacuum chamber and subjected to flood exposure by a 405 nm lamp (25.4 mW/cm$^2$ at sample location) for 5 minutes.

### 2.4 Photoresins used for fabrication

All parts printed in this work were made with a mixture of diurethane dimethacrylate (DUDMA) with poly(ethylene glycol) diacrylate (Mn=700 g/mol, PEGDA 700) in a 8:2 wt ratio. Ethyl (2,4,6-trimethylbenzoyl) phenylphosphinate (TPO-L) at a concentration of 1.75 mM was used as the photoinitiator. PEGDA 700 was purchased from Sigma Aldrich, DUDMA from EssTech Inc., and TPO-L from Oakwood Chemical. Prepared photoresins were transferred to 8 mm outer diameter vials (Fisher Scientific) and stored in a dark container at room temperature until air bubbles were removed from suspension. The refractive index at 405 nm of the liquid and fully-cured photoresin was measured to be 1.4997 and 1.5051 respectively using a Schmidt-Haensch ATR-BR refractometer. The room temperature viscosity of the photoresin was measured to be 1100 cP using a Brookfield DV-III Ultra Programmable Rheometer.

## 2.5 Evaluation of Printed Lens Imaging Performance

An example of a VAM- printed plano-convex lens (radius of curvature 3.1 mm, center thickness 1.5 mm, diameter 3 mm) is shown in Fig. 2. Here, the printed lens (Fig. 2(a)) was evaluated using a system of similar design of Ristok et al. [7]. As shown in Fig. 2(b), a white LED (Thorlabs MWUVL1) was collimated by a f = 30 mm plano-convex lens (Thorlabs LA1805) and filtered by a 670 nm bandpass filter (Thorlabs FBH670-10). The filtered and collimated beam was then transmitted through a ground glass diffuser (Thorlabs DG10-220-MD). A microscope objective (Olympus PLN 4X) functioned as a condenser, focusing the illumination light onto the VAM printed lens. Between the condenser and lens is a USAF 1951-1X negative MTF resolution pattern (Thorlabs R1DS1N). The resolution pattern as imaged by the VAM printed lens was captured by a second objective lens (Olympus PLN 10X) and reimaged onto a bare CMOS sensor (FLIR GS3-U3-91S6M) via a f = 100 mm plano-convex tube (Thorlabs LA1229) lens. Between the condensor and glass diffuser was an adjustable aperture that was used to set the illumination diameter on the lens to 2 mm.

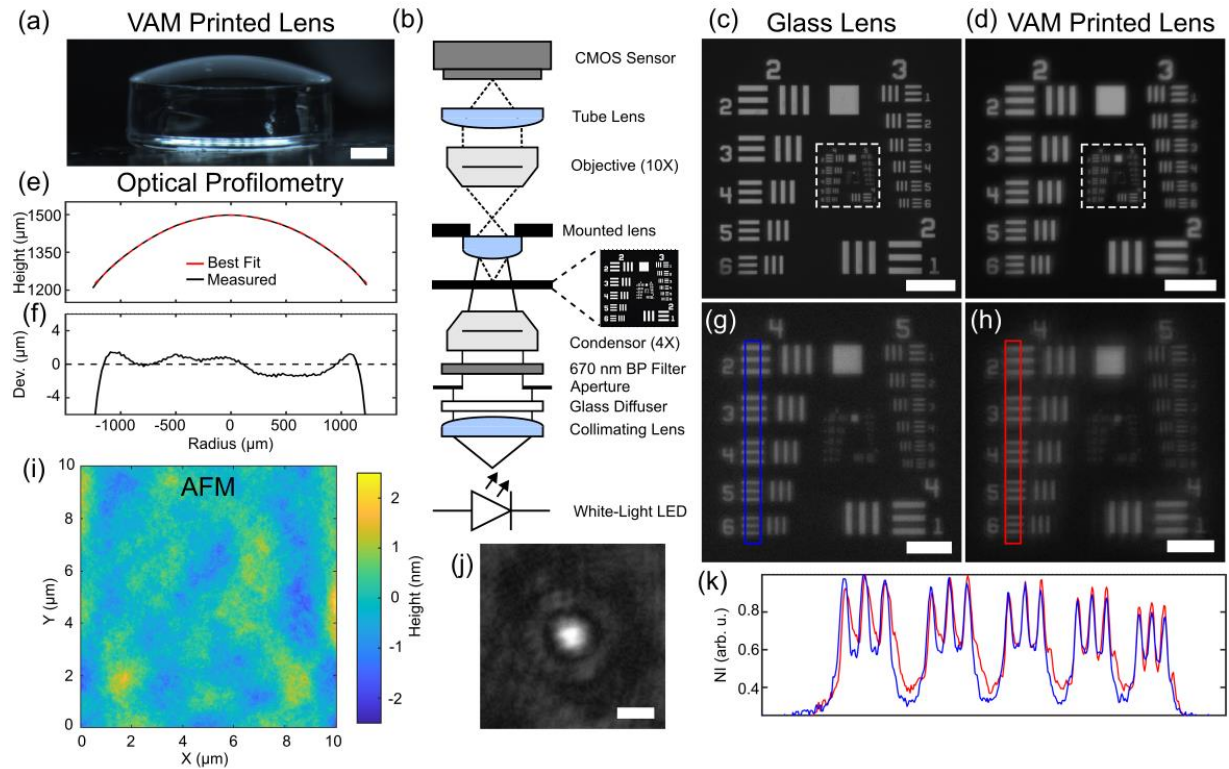

*Figure 2: Image performance comparison of glass and 3D printed lenses. (a) Image of printed lens. (b) Image resolution measurement device. Images of a USAF 1951-1X negative MTF resolution target as taken with the commercial glass lens (c) and printed lens (d). (e,f) Optical profilometry measurements of the convex surface of the printed lens. Magnified views of Groups 4 and 5 of the resolution target with commercial glass lens (g) and printed lens (h). (i) AFM measurement of the convex surface of the printed lens. (j) Image of 658 nm laser spot focused by printed lens. (k) Normalized intensity (NI) slices of the blue and red colored regions in (g,h). Scale bar in (a) is 500 µm. Scale bars in (c,d) are 250 µm  Scale bars in (g,h) are 62.5 µm. Scale bar in (j) is 10 µm.*

We compared the performance of our printed lens to a commercial glass lens of same dimensions (Thorlabs LA1036) Images of a negative USAF 1951 target acquired by both these lenses using the setup in Fig. 2(b) are shown in Fig. 2(c,g) and Fig. 2(d,h) respectively. Due to the difference in radius of curvature (2.94 mm vs 3.1 mm) and index of refraction between materials, images taken by the printed lens were magnified by 4% to match the scale of the glass lens. Qualitatively, the printed lens has comparable resolution to that of the commercial equivalent, indicating that the printed lenses here can achieve commercial-comparable imaging performance. The imaging resolution was quantified by taking a line profile through the Group 4 line pattern, and plotting the normalized intensity (NI) vs. field position as shown in Fig. 2(k) for the blue and red regions in Fig. 2(g) and 2(h) respectively. For all elements of Group 4 (17.96 lp/mm – 28.51 lp/mm), we observe comparable contrast for both lenses with values spanning 0.51 – 0.21 and 0.48 – 0.16 for glass and printed lens respectively. The smallest resolvable resolution element for the VAM-printed lens has a linewidth of 9.84 µm (group 5, element 5), whereas the commercial glass only resolves one further resolution element, with linewidth 8.77 µm (group 5, element 6), indicating that this VAM-printed lens reaches near commercial grade imaging performance. To underline this point further, we measured the spot size of a 658 nm laser that was focused by the printed lens. As shown in Fig. 2(j), we observe Airy-ring like features, with an least-squares fitted Gaussian full-width half maximum (FWHM) of $6.0 \pm 0.4$ µm which is close to the FWHM measured for the glass lens ($3.0 \pm 0.1$ µm).

We used optical profilometry and atomic force microscopy (AFM) to measure the shape error and root-mean-square (RMS) surface roughness of the printed lenses, respectively. As shown in Fig 2I, the measured height of the lens is close to the curvature of the glass lens (3.1 mm) with a least-squares best fit of $2.94 \pm 0.01$ mm. In Fig. 2(f) is given the deviation of the measured surface to the surface of best fit. We observe micron-scale deviation, with a RMS surface deviation of 1.4 µm over the 2.5 mm measurement window. In Fig. 2(i) is shown AFM measurements of a 10 µm X 10 µm area taken at the top of the convex lens surface. We obtain an RMS surface roughness of 0.53 nm, with peak values reaching + 2.73 nm/-2.14 nm. Furthermore, using optical microscopy we measured the diameter as 2.81 mm and is slightly smaller than nominal (3.00 mm).

## 2.6 Elimination of Striations and Printing of Microlens Arrays

Next, we used blurred tomography to print a 3x3 biconvex microlens array (MLA), with the model design shown in Fig. 3a. Here, each lenslet has a spherical surface with radius of curvature of 1 mm (effective focal length = 1.57 mm) and fills a 1 mm X 1 mm square aperture. Due to the size constraints of the printed part, the surface normal of the lenses were printed normal to the vial axis (Fig 3(a)). In VAM, this configuration results in layer-like defects called striations which manifest as ridges on the surface of the part that are parallel to the optical axis of the projector [10,11,12]. Striations can be seen in Fig 3(b) where the microlens array was printed using conventional VAM optical conditions [19]. Here, the etendue of the printer was set to 0.650 sr mm$^2$ and tomographic projections were optimized using ray tracing for only the chief ray (analogous to the Radon-based approach). As a result, imaging of a business card through the MLA is not possible as shown in the bottom-right pane of Fig. 3(b) where we observe strong blurring perpendicular to the striation direction. We investigated this further by examining the point-spread function (PSF) of the MLA after focusing of 658 nm laser illumination (experimental setup shown in Fig. 3(d)). The PSF was imaged onto a CMOS sensor by a 100 mm tube lens and a microscope objective (Olympus PLN 4X). As shown in Fig. 3(e), we observe strong elongation of the PSF perpendicular to the striation direction which is in agreement to the imaging experiment shown in Fig. 3(b).

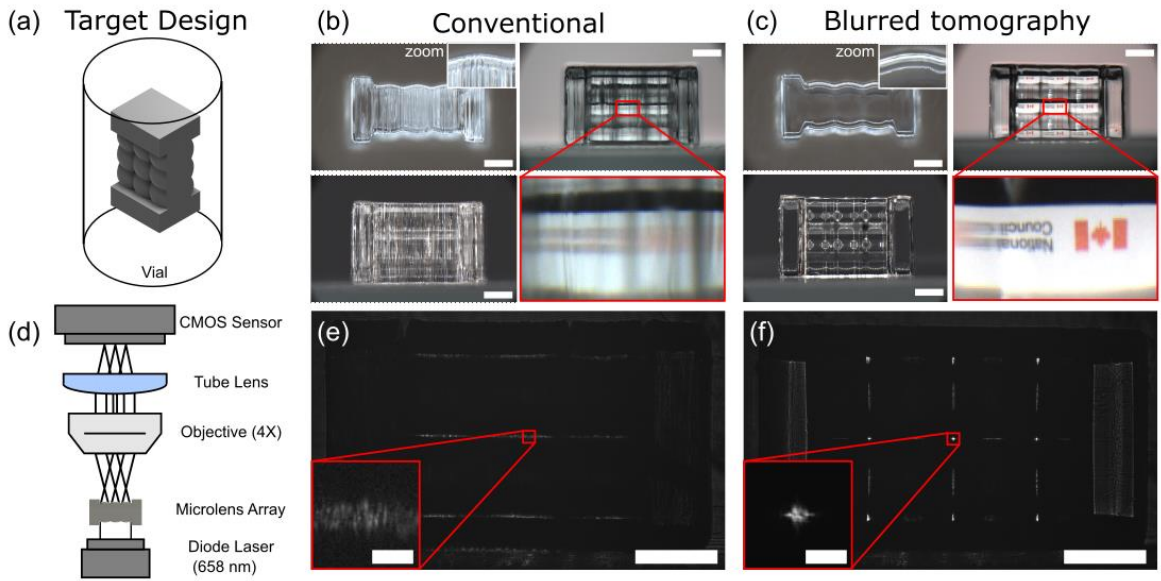

*Figure 3: Printing microlens arrays using blurred tomography. (a) Computer model of MLA inside vial (not to scale). Images of MLA printed using (b) conventional printing conditions and (c) blurred tomography. (d) Diagram of setup used to measure laser*

In contrast, the MLA printed with blurred tomography in Fig 3(c) does not exhibit striation defects since they are suppressed by the optical blurring imparted by the large etendue printing conditions. The removal of striations on imaging quality is evident in the right two panes of Fig. 3(c). All nine lenslets are able to image a business card placed behind the MLA, with lettering clearly visible in the zoomed-in image of the center lenslet. Furthermore, we measure a Gaussian spot size of 20 μm for the central lenslet as shown in Fig. 3(f). This distinguishes blurred tomography from previous reports of optics fabrication with VAM [12] since this method enables printing of parts with optical surfaces oriented normal to the writing beam.

*2.7 Optical Fiber Overprinting*

An advantage of VAM over other additive manufacturing techniques is the ability to print a 3D object over an existing object. In VAM this is referred to as overprinting and has been used to do print onto metallic substrates [11]. We have applied blurred tomography to print a ball lens onto an optical fiber. Ball lenses are commonly used to improve signal coupling between fiber optic components, such as focusing collimated light into a fiber, or conversely collimating light from a fiber. In Fig. 4 we demonstrate the latter. In Fig. 4(a) and 4(b) we show the target design and printed part. Here, the ball lens (diameter = 3 mm) is printed with a supporting structure to enable overprinting to the optical fiber (PMMA, diameter = 0.5 mm, NA = 0.23). The distance between the ball lens and fiber is given by the back focal length of the ball lens (0.84 mm at 658 nm) [21]. During printing, the optical fiber was aligned coaxially with the vial using a 3D-printed holder shown in Fig. 4(c). In Fig. 4(d) we show the experimental setup used to demonstrate collimation of fiber light by the printed ball lens. Here, a 658 nm diode laser was focused onto the bare end of the fiber by a 100 mm focal lens, and then the spot size of the beam after collimation by the printed ball lens was measured with a CMOS sensor. In Fig. 4(e) are images of the spot size with the CMOS sensor placed at three distances away from the printed lens. In comparison to the bare optical fiber (Supplemental S1) we see a reduction in beam divergence (5.4 degrees vs 13.5 degrees for the fiber).

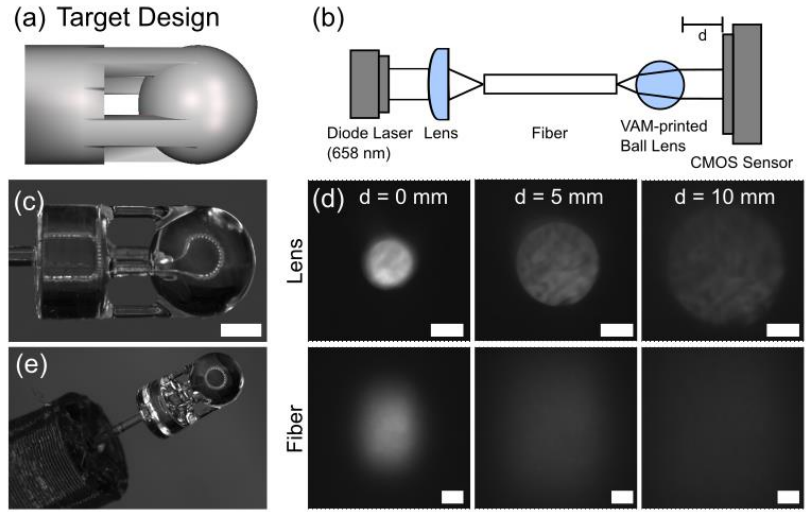

*Figure 4. Overprint of a ball lens onto an optical fiber. (a) Target design of ball lens. (b) Diagram of experiment used to measure collimating properties of printed ball lens. (c) Side view of printed ball lens on fiber. (d) Images of spot size of printed lens and bare fiber as measured by CMOS sensor located at different distances d from the end of the printed lens or bare fiber. (e) Ball lens and fiber mounted in 3D printed holder. All scale bars are 1 mm.*

*3 Discussion and Outlook*

We have demonstrated a new method to rapidly produce lenses with commercial-level quality using tomographic additive manufacturing. We call this method blurred tomography, and by purposely adding optical blurring to the writing beams, we are able to produce optically-smooth surfaces with good conformity to the target geometry. Blurred tomography enables low-cost and direct fabrication of optical components with the freeform design nature inherent to additive manufacturing techniques. Here, we have demonstrated fabrication of three different optical components: 1. A plano-convex singlet lens with sub-nm surface roughness (RMS = 0.53 nm), micron-scale surface deviation (RMS

= 1.4 μm), and commercially-comparable imaging resolution. 2. A biconvex microlens array, printed in an orientation previously not possible, and capable of imaging a business card, and 3. Overprinting of a ball lens onto a free-standing fiber optical fiber.

Typically in tomographic printing applications, the etendue of the writing beams must be small to achieve high resolution [10] and satisfy the requirements of the Filtered Backprojection method used to compute the tomographic projections [15]. This requirement is a double-edge sword as it also introduces layer-like artifacts into tomographic prints due to the self-writing waveguide effect [13]. Here, we have demonstrated that by purposely introducing optical blurring via increasing the etendue that we can alleviate these striations leading to optically smooth surfaces. The direct-printing nature of blurred tomography eliminates the need for special post-processing treatments used in VAM to create smooth surfaces (e.g. latent imaging, meniscus coating, sintering), making it simpler and more robust to implement. Furthermore, we have shown that the blurred tomography can achieve smooth surfaces for surfaces oriented normal to the writing beams (i.e. perpendicular to the vial axis), in contrast to previous reports that have only demonstrated fabrication of parts with optical surfaces oriented parallel to the vial axis [12,14]. Since blurred tomography is insensitive to part orientation, this paves the way to fabrication of larger optical components that utilize more of the cylindrical print volume, such as helical VAM [23]. It is important to note that in our work we have demonstrated a single optical configuration to achieve the necessary optical blurring. However, the method could be extended to other configurations such as index-matching bath [10,11,12,13,14] where optical astigmatism is not present, enabling more uniform blurring across the print volume.

The volumetric nature of blurred tomography makes it inherently faster to fabricate a part compared to other additive manufacturing techniques. For example, all lenses shown in this work were printed in < 1 minute, and completed post-processed and ready for handling within 30 minutes. These printing times are comparable to recent optics fabrication using VAM [12]. However, it is important to note that in [12] the part surface is not fully-cured and susceptible to damage, whereas parts made with our approach are fully cured and able to be handled. In another report but using 2PP [7] it had taken 23 hours to print a half-ball lens of comparable diameter (2 mm vs. 3 mm), with comparable radius of curvature error (3.9% vs 5.4%) and surface roughness (2.9 nm vs. 0.53 nm). This demonstrates that blurred tomography can achieve similar performance to 2PP with features at the millimeter-size scale but at a fraction of the fabrication time. The ability to rapidly go from design to part in hand will be invaluable for optical design refinement as many iterative design and fabrication cycles can be complete in a short time.

We anticipate this method to provide value for applications outside optical component fabrication as well. In particular, VAM is suitable for fabrication of hollow structures such as microfluidic channels [14,16,22] and printed arteries [10] as it benefits from the self-supporting nature of the print method. Here, blurred tomography could be used to produce smoother channels, reducing the potential for turbulent flow.

Supplementary

*S1: Measurement of Divergence Angle from Optical Fiber*

The divergence angle of the optical fiber with and without printed ball lens was measured using the experimental setup shown in Fig. 4(b). Here, a 658 nm laser was focused into one end of the optical fiber, and the emitted beam from the other end was measured using a CMOS camera sensor at different distances (Fig. S1(a)). The FWHM of the spot was measured by taking a central slice of the spot image, and fitting the slice to a 1D Gaussian (Fig. S1(b)). Next, the divergence angle of the beam, alpha, was determined by first fitting the measured FWHM vs. camera-fiber distance d (SFig. S1(c)) according to equation $\frac{FWHM}{2} = b_1(d + b_2)$, where $b_1$ and $b_2$ are fitting parameters. Next, the divergence angle $\alpha$ was determined via equation $\alpha = \tan^{-1}(b_1)$. For the bare optical fiber, the NA was determined via $NA = \sin(alpha)$.

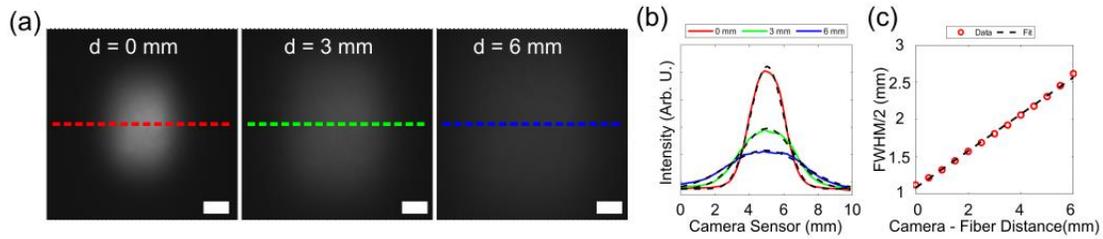

*Figure S1: Measurement of divergence angle of optical fiber. (a) Images taken with CMOS camera sensor at three different distances from fiber end. (b) 1D slices taken through camera sensor images with corresponding least-squares fit to a 1D Gaussian. (c) Fitted FWHM of Gaussian at different camera-fiber distances.*

**Funding**

Funding provided by the National Research Council Canada.

**Disclosures**

The authors declare the following financial interests/personal relationships which may be considered as potential competing interests: Daniel Webber, Antony Orth, Yujie Zhang, Michel Picard, Chantal Paquet, Jonathan Boisvert, have a patent pending to National Research Council Canada.

**Data Availability**

Data available upon reasonable request.

**Acknowledgements**

The authors would like to thank Liliana Gaburici for performing AFM measurements of the samples. The authors would also like to thank the anonymous reviewers for their time.